\documentclass[12pt]{article}
\usepackage{latexsym}
\usepackage{amsmath,amsbsy,amssymb}
\usepackage{verbatim}
\usepackage{bm}
\usepackage{cite}

\usepackage{curves}
\usepackage{epic}
\usepackage{eepic}
\usepackage{epsfig}

\newcommand{\dd}{\mbox{\rm d}}

\newcommand{\dg}{\dagger}

\newcommand{\tl}{\tilde}

\newcommand{\p}{\partial}
 
\newcommand{\be}{\begin{equation}}
\newcommand{\bear}{\begin{eqnarray}}
\newcommand{\ear}{\end{eqnarray}}
\newcommand{\ee}{\end{equation}}

\newcommand{\lbl}{\label}
\newcommand{\bi}{\bibitem}
\newcommand{\ci}{\cite}

\newcommand{\vs}{\vspace}
\newcommand{\hs}{\hspace}

\newcommand{\vphi}{\varphi}
\newcommand{\bbp}{{\bm p}}
\newcommand{\bp}{{\bar p}}

\newcommand{\bbx}{{\bm x}}
\newcommand{\bx}{{\bar x}}
\newcommand{\bpp}{{\bar \p}}

\newcommand{\vac}{|0 \rangle}

\newcommand{\sg}{\sigma}

\newcommand{\om}{\omega}

\begin{document}

\begin{center}

\

\vs{.8cm}

\baselineskip .7cm

{\bf \Large  Manifestly Covariant Canonical Quantization of the Scalar Field
and Particle Localization} 

\vs{4mm}

\baselineskip .5cm
Matej Pav\v si\v c

Jo\v zef Stefan Institute, Jamova 39,
1000 Ljubljana, Slovenia

e-mail: matej.pavsic@ijs.si

\vs{3mm}

{\bf Abstract}
\end{center}

\baselineskip .45cm

{\footnotesize Particle localization within quantum field theory is revisited. Canonical quantization of a free scalar field theory is performed in a manifestly Lorentz covariant way with respect to an arbitrary 3-surface $\Sigma$, which is the simultaneity surface associated with the observer, whose proper time direction is orthogonal to $\Sigma$. Position on $\Sigma$ is determined by a 4-vector ${\bar x}^\mu$. The corresponding quantum position operator, formed in terms of the operators $a^\dagger ({\bar x})$, $a({\bar x})$, that create/annihilate  particles on $\Sigma$, has thus well behaved properties under Lorentz transformations.  A generic state is a superposition of the states, created with $a^\dagger ({\bar x})$,  the superposition coefficients forming  multiparticle wave packet profiles---wave functions, including a single particle wave function that satisfies the covariant generalization of the Foldy equation. The covariant center of mass operator is introduced and its expectation values in a generic multiparticle state calculated.

\vs{2mm}

}

\baselineskip .55cm

\section{Introduction}

Quantum gravity is a tough and long lasting problem of theoretical physics. One of the possible
approaches to its resolution starts from the assumption that gravity is an effective theory of
a more fundamental theory, such as that of strings or branes\,\ci{DuffBenchmarks}.  Within such
a framework we can consider our 4-dimensional spacetime as a 4-dimensional surface, a brane's
``world volume'' or ``world sheet'', embedded in a higher dimensional space\ci{PavsicTapia}; and what we need is
quantization of such an object. While quantization of a string is rather well understood, quantization
of a generic brane has remained an unfinished project, whose resolution would be a significant step towards quantum gravity. In Refs.\,\ci{PavsicFlatBrane} a brane was considered as a bunch of point
particles, each one being described by a scalar field. To form a brane, one needs an uncountable
infinite set of operators that create particles at given positions. For this purpose, instead of expanding
a scalar field in terms of the creation and annihilation operators  in momentum space, one expands
it in terms of the creation and annihilation operators in position space. Using an uncountable set
of position creation operators and a suitable local interaction amongst them, we can form
wave packets, whose expectation values satisfy the equations of motion of a classical
Nambu-Goto brane\,\ci{PavsicFlatBrane}.

In view of such a promising approach to quantization of branes, it would be desirable to revisit
the role of position space in quantum field theories. A vast literature on this topics\,\ci{NewtonWigner,Wightman,
Fong,Barat,Cirilo-Lombardo,Fleming1,Fleming2,Rosenstein,Eckstein,Ruijsenaars,Al-Hashimi,Horwitz,Valente,
Nikolic1,Nikolic2,Karpov1,Karpov2,Lusanna1,Lusanna3,Padmanabhan1,Padmanabhan2,Hoffmann}
gives the impression that no final generally accepted resolution of the problem has been
achieved. 

Amongst the important insights on certain pieces of the puzzle let me mention the paper
by Rosenstein and Horwitz\,\ci{Horwitz} in which a description of a free relativistic spinless
particle within the framework of quantum field theory has been provided. The relation
between the Newton-Wigner-Foldy wave function and the Klein-Gordon wave function
was considered. The fact that the former wave function has not simple transformation
properties under Lorentz transformations was not taken as an argument against the
acceptability in QFT of wave packets, localized in position space.
Another important insight, although not generally accepted, is provided by Fleming\,\ci{Fleming1,Fleming2},
who proposed that particle localization refers to a given simultaneity 3-surface (hypersurface),
embedded in spacetime.

In this paper, motivated by the above mentioned works\,\ci{Horwitz,Fleming1,Fleming2}, we explore
the role of position in the free scalar field quantum theory. We canonically quantize 
a classical real scalar field, satisfying the Klein-Gordon equation, by performing a
covariant split of spacetime into space and time. ``Space" is associated with an arbitrary
3-surface $\Sigma$, and ``time'' is associated with the 1-dimensional manifold, parametrized
by $s \in \mathbb{R}$, along a unit 4-vectors $n^\mu$, orthogonal to $\Sigma$.
Such a split is covariant, because the 4- vector $n^\mu$, which determines it, is
arbitrary. A scalar field can be written as a function of $s$ and a 4-vector $\bx^\mu$ that
denotes position on $\Sigma$. One can then define creation and annihilation operators
$a(s,\bx)$ and $a^\dg (s,\bx)$ as certain combinations of the field $\vphi(s,\bx)$ and
the canonical momentum $\Pi = \p \vphi(s,\bx)/\p s$. The obtained relations are
straightforward  generalizations of the creation/annihilation operators of the {\it real}
harmonic oscillator. A generic state $|\Psi \rangle$ is then defined as a linear combination of the Fock space
basis states generated by the action of $a^\dg (\bx) \equiv a^\dg (0,\bx)$ on the vacuum
annihilated by $a (\bx) \equiv a (0,\bx)$. The superposition coefficients are $s$-dependent,
complex valued multiparticle wave functions $f(s,\bx_1,\bx_2,...,\bx_r)$. In the free field case
each multiparticle mode evolves independently of the other modes.
From the Schr\"odinger equation $i \p |\Psi \rangle/\p s = H |\Psi \rangle$ it then follows that a
one particle mode $f(s,\bx)$ satisfies the covariant generalization of the Foldy
equation\,\ci{Foldy,Horwitz}, $i \p f/\p s = \sqrt{m^2 + \nabla^2} f$,
 which  holds for the positive frequency part of the Klein-Gordon equation.
Although we quantized a {\it real} scalar field, the wave function $f(s,\bx)$ is {\it complex} in general,
similar to complex expansion coefficients of a generic state of a real harmonic oscillator.

The absolute value of the wave function, $f^* f$, gives the probability density of finding
at (proper) time $s$ a particle at position $\bx$ on $\Sigma$. Computation of the wave packets,
described by $f(s,\bx)$, where it was taken $\bx^\mu = (0,\bbx)$, has been considered
in Refs.\,\ci{Rosenstein, Al-Hashimi,Cirilo-Lombardo,PavsicLocal}. It was pointed out\,\ci{PavsicLocal}
that the existence of the wave function $f (s,\bx)$ in the quantum theory of a scalar field is unavoidable.
Although the probability density $f^* f$, in the case when the wave packet width is smaller than the
Compton wavelength, $\sg < 1/m$, can leak outside the light cone, this does not automatically
imply causal paradoxes\,\ci{Fleming1,Fleming2,PavsicLocal}. Anyway, if $\sg > 1/m$, no ``acausal"
behavior of $f(s,\bx)$ takes place\,\ci{Rosenstein,PavsicLocal}.

A covariant generalization of the Newton-Wigher position operators is defined as
${\hat \bx}^\mu = \int \dd \Sigma\, a^\dg (\bx) \, \bx^\mu \,a(bx)$;
its eigenstates are $a^\dg (\bx) \vac$ with the eigenvalues $\bx^\mu$ on $\Sigma$. The exact positions
are idealizations. In reality there is a spreading, determined by a wave packet $f(s,\bx^\mu)$.
And if $\sg < 1/m$, certain subtleties have to be taken into account:  in free case, a wave
packet after some time splits into two separate pieces, each  moving on average with the speed
of light\,\ci{Al-Hashimi,PavsicLocal}. But free case is an idealization, in reality sooner or
later an interaction takes place, and there is no longer a single particle state. Here
an interacting QFT is applicable. But it is important to fully understand the free case as well.

In Sec.\,2 we consider a covariant split of spacetime and introduce the operators that create and
annihilate particles at position $\bx^\mu$ on a hypersurface $\Sigma$, determined by an
arbitrary 4-vector $n^\mu$. In Sec.\,3 we discuss generic multiparticle state and the equations
of motion of the wave functions that determine them. We also define a covariant form of the position
operator, introduce the center of mass operators, and derive its expectation value
in a generic multiparticle state.

\section{Covariant split of spacetime in the scalar field theory and covariant position creation operators}

The scalar field operator, satisfying the Klein-Gordon equation that
can be derived from the action
\be
  I = \frac{1}{2} \int \dd^4 x \left ( \p_\mu \vphi \p ^\mu \vphi -m^2 \right ),
\lbl{2.1a}
\ee
is given by the manifestly Lorentz covariant expression
\be
  \vphi (x) = \int \frac{\dd^4 p}{(2 \pi)^4} c(p) \delta (p^2-m^2) {\rm e}^{-i p_\mu x^\mu},
\lbl{2.1}
\ee
where $x \equiv x^\mu$ and $p \equiv p^\mu$ are the position and momentum 4-vectors,
respectively.

Introducing a unit time like 4-vector $n^\mu$, normal to a 3-surface $\Sigma$,
and the projectior ${P^\mu}_\nu =
{\delta^\mu}_\nu - n^\mu n_\nu$, we can split the coordinates and momenta into
a normal part
\be
  x_{\bot} \equiv s = x^\mu n_\mu~,~~~~~~~~~~~~p_{\bot} = p_\mu n^\mu
\lbl{2.2}
\ee
and the orthogonal part
\be
  {\bar x}^\mu = {P^\mu}_\nu x^\nu~,~~~~~~{\bar p}_\mu = {P_\mu}^\nu p_\nu .
\lbl{2.3}
\ee
Similarly we can split the partial derivative\,\ci{BarutSplit,PavsicPointLikeAction}:
\be
  \p_\mu = (\p, {\bar p}_\mu)~,~~~~~\p \equiv n^\mu \p~,~~~~~~\bpp_\mu ={P_\mu}^\nu \p_\nu.
\lbl{2.3a}
\ee

In particular, the vector $n^\mu$ can coincide with the time like direction of the Lorentz
frame in which we work, so that $n^\mu = (1,0,0,0)$. We will consider a general case,
in which $n^\mu$ is arbitrary unite time like vector.

The volume elements $\dd^4 x$ and $\dd^4 p$ can then be factorized according to
\be
  \dd^4 x = \dd s \,\dd \Sigma~,~~~~ \dd^4 p = \dd p_{\bot} \dd \Sigma_\bp,
\lbl{2.4}
\ee
where
\be
  \dd \Sigma \equiv n^\mu \epsilon_{\mu \nu \rho \sigma} \dd \bx^\nu \dd \bx^\rho \dd \bx^\sg
~~~{\rm and}~~~~~\dd \Sigma_\bp 
   = n_\mu \epsilon^{\mu \nu \rho \sg} \dd \bp_\nu \bp_\rho \dd \bp_\sg
\lbl{2.5}
\ee
are the volume elements in the hypersurface $\Sigma$, orthogonal to $n_\mu$.

Using (\ref{2.3a}) and (\ref{2.4}), the action (\ref{2.1a}) can be expressed as
\be
  I = \frac{1}{2} \int \dd \Sigma \left ( \left (\frac{\p \vphi}{\p s}\right )^2
       +\bpp_\mu \bpp^\mu \vphi - m^2 \vphi^2 \right ) ,
\lbl{2.6a}
\ee
where $\p \vphi/\p s= n^\mu \p_\mu \vphi$. The pair of canonically
conjugated variables is $\vphi$ and $\Pi = \p \vphi/\p s$.

Using (\ref{2.2})--(\ref{2.5}), we can write Eq.\,(\ref{2.1}) in the form
\be
  \vphi(s,\bx) = \int \frac{\dd p_{\bot} \dd \Sigma_\bp}{(2 \pi)^4}\, c(p_{\bot},\bp)
  \delta (p_{\bot}^2 + \bp^\mu \bp_\mu - m^2) {\rm e}^{-i p x} ,
\lbl{2.6}
\ee
where $p x \equiv p_\mu x^\mu = p_{\bot} s +\bp_\mu \bx^\mu$ and $\bp \equiv \bp_\mu$.

Performing the integration over $\bp_\bot$, we obtain
\be
  \vphi (s,\bx) = \int \frac{\dd \Sigma_\bp}{(2 \pi)^3 2 \om_\bp} \left (  {\tl a}(\bp)
   {\rm e}^{-i p x} + {\tl a}^\dg (\bp) {\rm e}^{i p x}  \right ),
\lbl{2.7}
\ee
where ${\tl a} (\bp) = c(\om_\bp, \bp)/(2 \pi)$, ${\tl a}^\dg (\bp) = c(-\om_\bp,\bp)/(2 \pi)$,
and $\om_\bp = \sqrt{m^2-\bp_\mu \bp^\mu}$. Absorbing the factor $1/\sqrt{(2 \pi)^3 2 \om_\bp}$
into a redefinition of the operators ${\tl a}(\bp)$, ${\tl a}^\dg(\bp)$, we introduce
\be
 a(\bp) = \frac{{\tl a}(\bp)}{\sqrt{(2 \pi)^3 2 \om_\bp}}~,~~~~~
 a^\dg(\bp) = \frac{{\tl a}^\dg(\bp)}{\sqrt{(2 \pi)^3 2 \om_\bp}} ,
\lbl{2.7a}
\ee
so that
\be
  \vphi (s,\bx) = \int \frac{\dd \Sigma_\bp}{\sqrt{(2 \pi)^3 2 \om_\bp}} \left ( {a}(\bp)
   {\rm e}^{-i p x} + {a}^\dg (\bp) {\rm e}^{i p x} \right ).
\lbl{2.7b}
\ee

The quantity $\dd \Sigma_\bp$ is Lorentz invariant and so is $\om_\bp$. Therefore, $\vphi (x)$
as expressed in (\ref{2.7}) or (\ref{2.7b}) is manifestly Lorentz covariant. Regardless of the Lorentz frame
that we observe from, $\om_\bp$ is always the quantity defined with respect to the
chosen direction $n^\mu$ in spacetime.

The following commutation relations are satisfied\footnote
{In particular, if $n^\mu =(1,0,0,0)$, then $s=x^0 \equiv t$, $\bx^\mu = (0,x^i) \equiv (0,\bbp)$.}
\be
  [\vphi (s,\bx),\Pi (s,\bx')] = \delta^4 (\bx - \bx'),
\lbl{2.8a}
\ee
\be
  [\vphi (s,\bx),\vphi (s,\bx')] = 0~,~~~~~~~ [\Pi (s,\bx),\Pi (s,\bx')] = 0.
\lbl{2.8b}
\ee
The corresponding commutation relations for $a(\bp)$, $a^\dg (\bp)$ are
\be
  [a(\bp), a^\dg (\bp')] = \delta^4 (\bp-\bp') ,
\lbl{2.9a}
\ee
\be
   [a(\bp), a(\bp')=0~,~~~~~  [a^\dg(\bp), a^\dg (\bp')=0.
\lbl{2.9b}
\ee

The Hamilton operator, corresponding to the action (\ref{2.6a}), is
\be
  H=\frac{1}{2} \int \dd \Sigma \left ( \Pi^2 - \bpp_\mu \vphi \bpp^\mu \vphi + m^2 \vphi^2 \right ),
\lbl{2.10}
\ee
which after performing the partial integration and omitting the surface term (assumed to be zero),
becomes
\be
  H=\frac{1}{2} \int \dd \Sigma \left ( \Pi^2 + \vphi (m^2 + \bpp_\mu \bpp^\mu) \vphi \right ),
\lbl{2.11}
\ee

This is analogous to the Hamilton operator $H= \frac{1}{2} (p^2 + \om^2 q^2)$ of the harmonic
oscillator which, after introducing the operators $a = \frac{1}{\sqrt{2}} \left ( \sqrt{\om} q + \frac{i}{\sqrt{\om}} p \right )$,
$a^\dg = \frac{1}{\sqrt{2}} \left ( \sqrt{\om} q - \frac{i}{\sqrt{\om}} p \right )$, becomes $H= \om (a^\dg a + a a^\dg)$.
Because in the case of the Hamiltonian (\ref{2.11}) there are infinitely many variables 
$\vphi (s,\bx)$, $\Pi (s,\bx)$, distinguished by the label $\bx \equiv \bx^\mu$, let us now introduce
the operators\footnote{
The non covariant form of those operatore, as well as the Hamiltonian (\ref{2.11}), was considered by Jackiw\,\ci{Jackiw}.}
\be
  a(s,\bx) = \frac{1}{\sqrt{2}} \left ( \sqrt{\om_\bx} \vphi + \frac{i}{\sqrt{\om_\bx}} \Pi \right ),
\lbl{2.12}
\ee
\be
  	a^\dg(s,\bx) = \frac{1}{\sqrt{2}} \left ( \sqrt{\om_\bx} \vphi - \frac{i}{\sqrt{\om_\bx}} \Pi \right ),
\lbl{2.13}
\ee
where
\be
   \om_\bx \equiv \sqrt{m^2 + \bpp_\mu \bpp^\mu},
\lbl{2.13a}
\ee
satisfying
\be
  [a(s,\bx),a^\dg (s,\bx')] = \delta^4 (\bx-\bx'),
\lbl{2.14}
\ee
\be
  [a(s,\bx),a (s,\bx')] = 0~,~~~~~~[a(s,\bx),a (s,\bx')] = 0,
\lbl{2.15}
\ee
Using (\ref{2.12})--(\ref{2.15}), the Hamilton operator (\ref{2.11}) takes the form
\be
  H=\frac{1}{2} \int \dd \Sigma \left ( a^\dg (s,\bx) \om_\bx a (s,\bx) + a (s,\bx) \om_\bx a\dg (s,\bx) \right ).
\lbl{2.16}
\ee

If we insert into (\ref{2.12}), (\ref{2.13}) the expresioon (\ref{2.7b}) for $\vphi(s,\bx)$ and the
corresponding expression for $\Pi = \p \vphi/\p \bx$, then we obtain
\be
  a(s,\bx) = \frac{\sqrt{\om_\bx}}{\sqrt{2}} \int \frac{\dd \Sigma_\bp}{\sqrt{(2 \pi)^3 2 \om_\bp}}
  a(\bp) {\rm e}^{- i p x} .2 = \frac{1}{\sqrt{(2 \pi)^3}} \int \dd \Sigma_\bp a (\bp) {\rm e}^{-i p x} ,
\lbl{2.17a}
\ee
\be
  a^\dg(s,\bx) = \frac{\sqrt{\om_\bx}}{\sqrt{2}} \int \frac{\dd \Sigma_\bp}{\sqrt{(2 \pi)^3 2 \om_\bp}}
  a^\dg(\bp) {\rm e}^{ i p x} .2 = \frac{1}{\sqrt{(2 \pi)^3}} \int \dd \Sigma_\bp a^\dg (\bp) {\rm e}^{i p x} ,
\lbl{2.17b}
\ee
In the above calculations we used $p x \equiv p_\mu x^\mu = p_\bot x_\bot + \bp_\mu \bx^\mu$.

Written in terms of $a(\bp)$, $a^\dg (\bp)$, the Hamiltonian is
\be
  H= \frac{1}{2} \int \dd \Sigma_\bp \, \om_\bp \left ( a^\dg (\bp) a(\bp) + a(\bp) a^\dg (\bp) \right ).
\lbl{2.18}
\ee

We see that between the operators $a(\bp)$, $a^\dg (\bp)$ and $a(\bx)$, $a^\dg (\bx)$ there are
the covariant relations (\ref{2.17a}) and (\ref{2.17b}). The operators $a (\bp)$, $a^\dg (\bp)$ are
just the covariant forms of the usual operators $a(\bbp)$, $a^\dg (\bbp)$, which are generally accepted
as legitimate operators of QFT. On the contrary, the Fourier transformed operators
$a (\bbx) = \frac{1}{\sqrt{(2 \pi)^3}} \int \dd^3 \bbp \, a(\bbp) {\rm e}^{-i \bbp \bbx}$, whose Hermitian
conjugate $a^\dg (\bbx)$ creates the Newton-Wigner localized states are usually not considered
as meaningful operators\,\ci{WightmanSchweber,Teller,Padmanabhan1}, for various
reasons, mainly related to their Lorentz
(non) covariance properties. However, in our manifestly Lorentz covariant formulation of QFT,
the operators $a(\bbx)$, $a^\dg (\bbx)$ are a special case of the covariant operators $a(\bx)$,
$a^\dg (\bx)$, $\bx \equiv \bx^\mu = {P^\mu}_\nu x^\nu$. As $a(\bp)$, $a^\dg (\bp)$, also
$a (\bx)$, $a^\dg (\bx)$ are unavoidable ingredients of the quantum field theory. They cannot be
swept under the carpet.

Thus, introducing the vacuum $\vac$ such that $a(\bx) \vac = 0$, the operators $a^\dg (\bx)$ create
the states with definite position $\bx \equiv \bx^\mu$ on a hypersurface $\Sigma$:
\be
  |\bx \rangle = a^\dg (\bx) \vac .
\lbl{2.19}
\ee
Similarly, the operators $a^\dg (\bp)$, $\bp \equiv \bp^\mu$ create the particles with the momentum,
whose space like part, $\bp^\mu$, is in the hypersurface $\Sigma$.

A hypersurface $\Sigma$ is the simultaneity 3-surface of an observer ${\cal O}_n$, whose proper
time is associated with the direction $n^\mu$, orthogonal to $\Sigma$. The 4-vectors $\bx^\mu$
denote position on $\Sigma$. All those positions on $\Sigma$ are mutually simultaneous; they form
the 3-space, experienced by ${\cal O}_n$. Successive applications of the operators $a^\dg (\bx)$ on
the vacuum $\vac$ create a multiparticle state
\be
  |\bx_1,\bx_2,\bx_3,... \rangle = a^\dg (\bx_1) a^\dg (\bx_2) a^\dg (\bx_3) ... \vac ,
\lbl{2.20}
\ee
with positions on $\Sigma$.

Equations (\ref{2.19}) and (\ref{2.20}) denote Lorentz covariant description of position states.
It refers to the Lorentz reference frame in which positions on a given hypersurface $\Sigma$ are
given by the 4-vector $\bx^\mu$. In another Lorentz frame the positions on the same $\Sigma$
are given by the 4-vector $\bx'^\mu = {L^\mu}_\nu \bx^\nu$ (i.e., $\bx'= L \bx$), and Eq.\,(\ref{2.19})
reads $|\bx' \rangle = a'^\dg (\bx') \vac$, where $a'^\dg (\bx') = a^\dg (L^{-1} \bx')$.

That spatial positions refer to a given 3-surface (hypersurface) $\Sigma$ in spacetime, is incorporated
in the formulation of relativity. We talk about the observers,  simultaneity surface, length contraction, etc.,
where ``length'' refers to a distance within a given $\Sigma$, and ``contractions'' refers to how such
a length looks in another frame.

\section{Wave packet states}

Realistic states have neither exact definite positions nor momentum. They are the states
with certain multiparticle wave packet profiles:
$$
  |\Psi \rangle = \sum_{r=0}^N \int \dd \Sigma_{\bp_1} \dd \Sigma_{\bp_2} ... \dd \Sigma_{\bp_r}
  g(s,\bp_1,\bp_2,...,\bp_r ) a^\dg (\bp_1) a^\dg )\bp_2) ...a^\dg (\bp_r) \vac
$$
\be
  \hs{2.3cm} = \sum_{r=0}^N \int \dd \Sigma_{1} \dd \Sigma_{2} ... \dd \Sigma_{r}
  f(s,\bx_1,\bx_2,...,\bx_r ) a^\dg (\bx_1) a^\dg )\bx_2) ...a^\dg (\bx_r) \vac .
\lbl{3.1}
\ee
A particular case is a single particle state
\be
   |\Psi \rangle =  \int \dd \Sigma_{\bp} \, g(s,\bp ) a^\dg (\bp)  \vac
   = \int \dd \Sigma \, f(s,\bx ) a^\dg (\bx)  \vac
\lbl{3.2}
\ee
The relation between $g$ and $f$ is the Fourier transformation
\be
  f(s,\bx) = \frac{1}{\sqrt{(2 \pi)^3}} \int \dd \Sigma_\bp \, g(s,\bp) {\rm e}^{i \bp_\mu \bx^\mu} .
\lbl{3.3}
\ee

The same state $|\Psi \rangle$ of Eq.\,(\ref{3.2}) can as well be expressed in terms of the operators
${\tl a}^\dg (\bp) = \sqrt{(2 \pi)^3 2 \om_\bp} a^\dg (\bp)$ and the wave packet profiles
${\tl g} (s,\bp) = \sqrt{(2 \pi)^3 2 \om_\bp} g(s,\bp)$, or ${\tl a}(\bx) = (2 \om_\bx)^{-1/2} a^\dg (\bx)$
and ${\tl f}(s,\bx) = ( 2 \om_\bx)^{-1/2} f (s,\bx)$ as:
\be
  |\Psi \rangle = \int \frac{\dd \Sigma_\bp}{(2 \pi)^3 2 \om_\bp}\,{\tl g} (s,\bp) {\tl a}^\dg (\bp) \vac =
  \int \dd \Sigma \sqrt{2 \om_\bx}\, {\tl f} (s,\bx) \sqrt{2 \om_\bx}\, {\tl a}^\dg (\bx) \vac .
\lbl{3.2a}
\ee

A generic state $|\Psi \rangle$ satisfies the Schr\"odinger equation
\be
  i \frac{| \Psi \rangle}{\p s} = H |\Psi \rangle ,
\lbl{3.4}
\ee
with the Hamiltonian (\ref{2.16}) or (\ref{2.18}).

Inserting the multiparticle state (\ref{3.1}) into the Schr\"odinger equation (\ref{3.4}),
we obtain the following equations for the wave packet profiles:
\be
  i \frac{\p g(s,\bp_1,\bp2,...,\bp_r}{\p s})= \sum_{k=1}^r \om_{\bp_k} g(s,\bp_1,\bp_2,...,\bp_k,...\bp_r),
\lbl{3.5}
\ee
\be
  i \frac{\p f(s,\bx_1,\bx2,...,\bx_r}{\p s}=  \sum_{k=1}^r \om_{\bx_k} f(s,\bx_1,\bx_2,...,\bx_k,...,\bx_r),
\lbl{3.5a}
\ee
in which we omitted the zero point energy.

In the case of a single particle state the latter equations are
\be
i  \frac{\p g(s,\bp)}{\p s} = \om_\bp \,g(s,\bp),
\lbl{3.6}
\ee
\be
i  \frac{\p f(s,\bx)}{\p s} = \om_\bx f(s,\bx),
\lbl{3.7}
\ee

Solving Eq.\,(\ref{3.6}), we obtain
\be
   g(s,\bp) = {\rm e}^{-i \om_\bp s} g(\bp) .
\lbl{3.6a}
\ee
Notice that {\it only positive frequencies occur in the solution for} $g(s,\bp)$, because
Eq.\,(\ref{3.6}) has the first order derivative with respect to the evolution
parameter $s$. The wave function $f(s,\bx)$, related to $g(s,\bp)$ according to (\ref{3.3}),
is a positive frequency solution of the Klein-Gordon equation
\be
  \frac{\p^2 \phi}{\p s^2} + \bpp_\mu \bpp^\mu \phi + m^2 \phi = 0,
\lbl{3.6b}
\ee
that can be factorized into the set of two equations
\be
  i \frac{\p \phi^{(+)}}{\p s} = \sqrt{m^2 + \bpp_\mu \bpp^\mu}\,\phi^{(+)} ,
\lbl{3.6c}
\ee
\be
  i \frac{\p \phi^{(-)}}{\p s} = -\sqrt{m^2 + \bpp_\mu \bpp^\mu}\,\phi^{(-)} .
\lbl{3.6d}
\ee
Within the formalism of the QFT of a scalar field, a single particle wave packet
state is given by Eq.\,(\ref{3.2}), where $f(s,\bx) \equiv \phi^{(+)} (s,\bx)$
satisfies Eq.\,(\ref{3.7}), i.e., the positive frequency component (\ref{3.6c})
of the Klein-Gordon equation (\ref{3.6b}).

In relativistic quantum mechanics before the second quantization, the wave function
satisfies the Klein-Gordon equation or its two component equivalent (\ref{3.6c}),
(\ref{3.6d}). Both, positive and negative frequencies are present. Consideration
of only positive frequencies is taken as an ad hoc restriction. But if one considers the
wave function as a wave packet profile of a single particle state of the scalar
field QFT, one finds that it automatically contains only positive frequencies.
Negative frequencies are excluded, because the corresponding operators $a(\bp)$
or $a(\bx)$ are assumed to annihilate the vacuum.

Concentrating on single particle states we find that the scalar product is\footnote{
In particular, if $n^\mu =(1,0,0,0)$, this becomes
$$\langle \Psi|\Psi \rangle = \int \dd^3 \bbp\, g^* (t,\bbp) g(t,\bbp) =
\int \dd^3 \bbx \, f^* (t,\bbx) f(t,\bbx).$$}
\be
  \langle \Psi|\Psi \rangle = \int \dd \Sigma_\bp\, g^* (s,\bp) g(s,\bp) =
   \int \dd \Sigma \, f^* (s,\bx) f(s,\bx) .
\lbl{3.8}
\ee
Using (\ref{3.2a}), the scalar product can also be written as
\be
  \langle \Psi|\Psi \rangle = \int \frac{\dd \Sigma_\bp}{(2 \i)^3 2 \om_\bp}\, {\tl g}^* (s,\bp) {\tl g}(s,\bp) =
   \int \dd \Sigma \, \sqrt{2 \om_\bx} \,{\tl f}^* (s,\bx) \sqrt{2 \om_\bx}{\tl f} (s,\bx)
\nonumber
\ee\
\be
  =  \int \dd \Sigma \left ( (\om_\bx {\tl f}^*) {\tl f} + {\tl f}^* \om_\bx {\tl f} \right ) 
  = i \int \dd \Sigma\,\left ( {\tl f}^* \frac{\p {\tl f}}{\p s} - \frac{\p {\tl f}^*}{\p s} {\tl f} \right ) .
\lbl{3.10}
\ee
In the case $n^\mu = (1,0,0,0)$ this becomes the familiar scalar product of relativistic quantum
mechanics. But now the wave function ${\tl f}$ does not contain positive and negative
frequencies; it contains only positive frequencies, as a consequence of the fact that
$f \equiv \phi^{(+)}$ satisfies the positive frequency equation (\ref{3.6c}), and so also does
${\tl f} = (2 \om_\bx)^{-1/2} f$. However, the expressions under the integral,
namely $f^*f = \om_\bx^{1/2} {\tl f}^* \om_\bx^{1/2} {\tl f}$ and $(1/2)(\om_\bx {\tl f}^* {\tl f}
+{\tl f}^* \om_\bx f)$ are not the same. They differ by a total derivative term, a consequence
being that whilst $f^* f$ is always positive, $(1/2)(\om_\bx {\tl f}^* {\tl f}+{\tl f}^* \om_\bx f)$
can be negative even if ${\tl f}$ contains only positive frequencies\ci{Horwitz}.

The probability density is thus $\rho = f^* f$. Differentiating it with respect to $s$ and
using (\ref{3.7}), we obtain
\be
  \frac{\p \rho}{\p s} = \frac{\p f^*}{\p s} f + f^* \frac{\p f}{\p s} = i \left ( \om_\bx f^* f - f^* \om_\bx f \right )
   = \bpp^\mu {\bar j}_\mu ,
\lbl{3.10a}
\ee
where
\be
   {\bar j}_\mu = \frac{i}{2 m} \left ( \bpp_\mu f^* f - f^* \bpp_\mu f \right ) + {\rm higher~order~terms} ,
\lbl{3.10b}
\ee
The higher order terms in the expression for the current ${\bar j}_\mu$ come from the
expansion of $\om_\bx = \sqrt{m^2 + \bpp^\mu \bpp_\mu}$  $ = m \left ( 1 + \frac{\bpp^\mu \bpp_\mu}{2 m^2} + ..., \right )$.

In this covariant formalism the probability density $\rho$ is Lorentz scalar, whilst the current ${\bar j}_\mu$ is
4-vector. There is no need to combine $\rho$ and ${\bar j}_\mu$ so to form components of a 4-vector.

A state with a definite position
\be
  |\bx_0^\mu \rangle = a^\dg (\bx_0) \vac
\lbl{3.11}
\ee
is a special case of the generic state (\ref{3.2}) for
\be
  f(0,\bx) = \delta^4 (\bx - \bx_0).
\lbl{3.12}
\ee
This is a covariant form of the Newton-Wigner localized state. The same state can as well be
expressed according to (\ref{3.2a}) with
\be
  {\tl f} (0,\bx) = (2 \om_\bx)^{-1/2} \delta^4 (\bx - \bx_0) ,
\lbl{3.13}
\ee
and
\be
  {\tl a}^\dg (\bx) = (2 \om_\bx)^{-1/2} a^\dg (\bx).
\lbl{3.14}
\ee
Because all those expressions are covariant there is no ambiguity of how they
transform under Lorentz transformations. Only when fixing $n^\mu$, the expressions
lose their manifest covariance. For instance, if observed from the reference frame
in which $n^\mu = (1,0,0,0)$ those expressions assume the non covariant forms:
\be
  |\bx_0 \rangle = |\bbx_0 \rangle = a^\dg (\bbx_0) \vac = (2 \om_\bx)^{-1/2}|_{\bbx=\bbx_0}
  a^\dg (\bbx) \vac,
\lbl{3.15}
\ee
\be
   f(0,\bx) = f(0,\bbx) = \delta^3 (\bbx-\bbx_0)~,
   ~~~~~{\tl f} (,\bbx) = (2 \om_\bx)^{-1/2} \delta^3 (\bbx-\bbx_)).
\lbl{3.16}
\ee

The states (\ref{3.11}) are eigenstates of the operator\footnote{
Expressed in terms of ${\tl a} (\bx)$, ${\tl a}^\dg (\bx)$ we have
$${\hat \bx}^\mu = \int \dd \Sigma \sqrt{2 \om_\bx} \,{\tl a}^\dg (\bx) \bx^\mu \sqrt{2 \om_\bx} \,{\tl a} (\bx)=
\int \dd \Sigma  \,{\tl a}^\dg (\bx) \sqrt{2 \om_\bx}\left ( \bx^\mu \sqrt{2 \om_\bx} \,{\tl a} (\bx) \right ),$$
which after the evaluation of the action of the operator $\om_\bx$ on the expression within the bracket becomes the covariant
form of the cumbersome expression, usually known as the Newto-Wigner position operator.}
\be
  {\hat \bx}^\mu = \int \dd \Sigma \, a^\dg (\bx) \bx^\mu a (\bx),
\lbl{3.17}
\ee
satisfying  ${\hat \bx}^\mu | \bx \rangle = \bx^\mu |\bx \rangle$. This is the covariant form of the Newton-Wigner position operator
\be
  {\hat \bbx} = \int \dd^3 \bbx \, a^\dg (\bbx) \bbx \,a(\bbx),
\lbl{3.18}
\ee
which is a particular case of (\ref{3.17}) for $n^\mu = (1,0,0,0)$. One can verify that
\be
  [{\hat \bx}^\mu, {\hat \bx}^\nu ] =0.
\lbl{3.18a}
\ee

Similarly we can define the momentum operator
\be
   {\hat \bp}^\mu = \int \dd \Sigma_\bp \, a^\dg (\bp) \bp^\mu \,a (\bp) =
   \int \dd \Sigma \,a^\dg (\bx) (- i \bpp_\mu) a (\bx),
\lbl{3.19}
\ee
whose eigenstates are $|\bp^\mu \rangle = a^\dg (\bp) \vac$.

The commutation relations between the position and momentum operator, defined according
to (\ref{3.17}) and (\ref{3.19}), are
\be
  [{\hat \bx}^\mu, {\hat \bp}_\nu ] = i {\hat N} {\delta^\mu}_\nu ,
\lbl{3.20}
\ee
where
\be
  {\hat N} = \int \dd \Sigma\, a^\dg (\bx) a(\bx) = \int \dd \Sigma_\bp \,a^\dg (\bp) a(\bp)
\lbl{3.21}
\ee
is the particle number operator.

If instead of a single particle state we consider a multiparticle state 
$\left ( \prod_{n=1}^N a^\dg (\bx_n) \right ) \vac$ and act on it by the position operator
(\ref{3.17}) we obtain
\be
  {\hat \bx}^\mu \left ( \prod_{n=1}^N a^\dg (\bx_n) \right ) \vac=
  \left ( \sum_{n=1}^N \bx^\mu \right ) \left ( \prod_{n=1}^N a^\dg (\bx_n) \right ) \vac .
\lbl{3.22}
\ee
Using (\ref{3.21}), we can define {\it the center of mass position operator}
\be
  {\hat \bx}_{\rm T}^\mu = {\hat N}^{-1} {\hat \bx}^\mu ,
\lbl{3.23}
\ee
which gives
\be
   {\hat \bx}_{\rm T}^\mu \left ( \prod_{n=1}^N a^\dg (\bx_n) \right ) \vac
   = \bx_{\rm T}^\mu \left ( \prod_{n=1}^N a^\dg (\bx_n) \right ) \vac ,
\lbl{3.24}
\ee
where
\be
  \bx_{\rm T}^\mu = \frac{1}{N} \sum_{n=1}^N \bx_n^\mu 
\lbl{3.25}
\ee
is the center of mass position. Its commutator with the momentum operator  is
\be
  [{\hat \bx}_{\rm T}^\mu , {\hat \bp}^\nu ] = i {\delta^\mu}_\nu ,
\lbl{3.26}
\ee
while
\be
  [{\hat \bx}_{\rm T},{\hat \bx}_{\rm T}^\nu ] = 0.
\lbl{3.26}
\ee
The latter commutator can be derived by using Eqs.\,(\ref{3.23}),(\ref{3.18a}),(\ref{3.20}) and
\be
  [{\hat \bx}^\mu, {\hat N} ] = 0~,~~~~~[{\hat \bx}^\mu,1] = [{\hat \bx}^\mu,{\hat N}^{-1} {\hat N}]
 = {\hat N}^{-1}  [{\hat \bx}^\mu, {\hat N} ] +  [{\hat \bx}^\mu, {\hat N}^{-1} ] {\hat N} = 0 ,
\lbl{3.28}
\ee
from which it follow $ [{\hat \bx}^\mu, {\hat N}^{-1} ] = 0$. 

The above equations (\ref{3.17})--(\ref{3.28}) are covariant generalizations of the equations
considered in Refs.\,\ci{PavsicLocal,PavsicFlatBrane}.

Expectation value of the center of mass position operator in an $N$-particle state is
\be
  \langle \Psi|{\hat \bx}_{\rm T}^\mu |\Psi \rangle =
  \int \dd \Sigma_1 \dd \Sigma_2 ... \dd \Sigma_N\,f^* (\bx_1,\bx_2,...,\bx_N) \bx_{\rm T}^\mu
  f(\bx_1,\bx_2,...,\bx_N) .
\lbl{3.29}
\ee
In the case of a single particle state the above equation becomes
\be
   \langle \Psi|{\hat \bx}_{\rm T}^\mu |\Psi \rangle = \langle \Psi|{\hat \bx}^\mu |\Psi \rangle =
   \int \dd \Sigma \, f^*(\bx) \bx^\mu f (\bx).
\lbl{3.30}
\ee

For the expectation value of the momentum operator we obtain
\be
  \langle \Psi|{\hat \bp}^\mu |\Psi \rangle = \int \dd \Sigma_{\bp_1} \dd \Sigma_{\bp_2}...\dd \Sigma_{\bp_N}
  \,g^* (\bp_1,\bp_2,...,\bp_N) \left (\sum_{n=1}^N \bp_n \right ) g(\bp_1,\bp_2,...,\bp_N) .
\lbl{3.31}
\ee

The problem of an appropriate definition of the relativistic center of mass has a long history and has
no unambiguous or unique solution\,\ci{Price}. For a recent insight see Ref.\,\ci{Lusanna2,Lusanna3,Lusanna4}.
Within the framework of the quantized scalar field in {\it Minkowski space} the center of mass operator is defined according
to (\ref{3.23}) in terms of the position operators on a given 3-surface $\Sigma$ in spacetime.
Such operator has well defined properties under Lorentz transformations, and its expectation value
is given by Eq.\,(\ref{3.29}). The center of mass of identical bosons is thus defined with respect
to a given 3-surface $\Sigma$, which is the simultaneity 3-surface of a given observer.
Within our formalism spatial position and the center of mass are represented as 4-vectors,
$\bx^\mu$ and $\bx_{\rm T}^\mu$, restricted to $\Sigma$. Observed from another frame, those
4-vectors become $\bx'^\mu = {L^\mu}_\nu \bx^\nu$, ${\bx'^\mu}_{\rm T} = {L^\mu}_\nu \bx_{\rm T}^\nu$,
and they still refer to the same 3-surface $\Sigma$. 

\section{Discussion and conclusion}

In the canonical quantization of a classical field, satisfying a relativistic wave equation,
such as the Klein-Gordon equation, one usually splits spacetime into time and space
according to $x^\mu = (x^0,x^i) \equiv (t,\bbx)$, $i=1,2,3)$. Therefore, the procedure lacks
manifest Lorentz covariance from the very beginning. A consequence, already observed
by Fleming\,\ci{Fleming1,Fleming2}, is a confusion and ambiguity about the meaning of spatial position and
the corresponding position operator. In this paper we have performed a covariant split of spacetime
into a 3-surface (hypersurface) $\Sigma$ and a 1-dimensional manifold, parametrized by $s$,
along the orthogonal direction to $\Sigma$,
defined by a unit 4-vector $n^\mu$. All quantities in our procedure are thus defined in terms of $s$
and the 4-vectors, restricted to lie within $\Sigma$, which is the simultaneity 3-surface
(``space'') of a given observer ${\cal O}_n$. A field operator is thus expressed in terms
of the operators $a (\bp)$, $a^\dg (\bp)$ that annihilate or create a particle with the momentum,
whose space like part, $\bp \equiv \bp^\mu$, is on $\Sigma$. Such operators can,
of course, be Fourier transformed into the operators $a(\bx)$, $a^\dg (\bx)$ that annihilate or
create a particle at the position $\bx \equiv \bx^\mu$ on $\Sigma$.  The occurrence of
the operators $a(\bx)$, $a^\dg (\bx)$ and the corresponding position operator
${\hat \bx}^\mu = \int \dd \Sigma \, a^\dg (\bx) \bx^\mu \, a (\bx)$ is unavoidable in the
scalar field theory: once we have $a (\bp)$ and $a^\dg (\bp)$, we automatically also have
$a(\bx)$ and $a^\dg (\bx)$.

Spatial position and position operators are thus defined with respect to $\Sigma$. The
eigenvalues of the so defined position operator are the positions as observed by the
observer ${\cal O}_n$. In our covariant formalism we describe such positions not within
the reference frame of the observer ${\cal O}_n$, in which $n^\mu=(1,0,0,0)$, but within
an arbitrary reference frame.
Because position, despite being restricted to $\Sigma$, is defined
as a 4-vector $\bx^\mu$ in spacetime, it has well behaved properties with respect to
Lorentz transformations. In another Lorentz frame it is still position on the same 3-surface
$\Sigma$, but given in terms of the transformed coordinates $\bx'^\mu = {L^\mu}_\nu \bx^\nu$.

The meaning of position and position operator in the so formulated scalar field theory is thus
unambiguous, and the existence of the operator itself, unavoidable. It has no strange properties
under Lorentz transformations.

By using the position or momentum creation operators one can form multiparticle wave packed
states, the superposition coefficients being multiparticle wave functions
$f(s,\bx_1,\bx_2,...,\bx_N)$ or $g(s,\bp_1,\bp_2,...,\bp_N)$. In the case of a single particle
state, the wave functions $f(s,\bx)$ and $g(s,\bp)$ satisfy, respectively, the covariant
generalizations of the Foldy equation\,\ci{Foldy,Horwitz}, $i \p f/\p s = \om_\bx f$ and
$i \p g/\p s = \om_\bp g$, which involves positive frequencies only. Solutions for such
relativistic wave functions have been considered in Refs.\,\ci{Al-Hashimi,PavsicLocal}.
Whether or not such a wave functions violates causality is a subject of discussion. While
the authors, following Hegerfeldt's arguments concerning the amount of the single particle probability density
outside the light cone\,\ci{Hegerfeldt,Hegerfeldt2}, assert that
causality is violated,  others\,\ci{Fleming1,Fleming2,Ruijsenaars,Valente,Karpov1,Karpov2,Wagner,Hoffmann}
have argued that a paradoxical detection of a particle from the future cannot occur by such means.
In Ref.\,\ci{PavsicLocal} it was pointed out that what is crucial for causality issues is not merely
a detection of a particle from the  ``future'', but detection of a {\it signal} from the future. The transmission of information by sending signals from the future into the
past cannot be achieved at the macroscopic lever with the relativistic wave packets, even if they spread
outside the light cone. Therefore, no causal paradoxes of the ``grand father type" can
occur with such wave packets.

\end{document}